\documentclass[%
aip,
amsmath,amssymb,
reprint,%
]{revtex4-1}

\usepackage{graphicx}
\usepackage{dcolumn}
\usepackage{bm}

\usepackage[utf8]{inputenc}
\usepackage[T1]{fontenc}
\usepackage{mathptmx}
\usepackage{etoolbox}

\usepackage[dvipsnames]{xcolor}

\makeatletter
\def\@email#1#2{%
 \endgroup
 \patchcmd{\titleblock@produce}
{\frontmatter@RRAPformat} {\frontmatter@RRAPformat{\produce@RRAP{*#1\href{mailto:#2}{#2}}}\frontmatter@RRAPformat} {}{} }%
\makeatother

\newcommand{\reffig}[1]{Fig.~\ref{#1}}
\newcommand{\kT}{k_{\mathrm{B}}T}

\newcommand{\done}[1]{\colorbox{green}{\textbf{DONE}}}
\newcommand{\doneDM}[1]{\colorbox{green}{\textbf{DONE by DM}}}

\begin{document}
\preprint{AIP/123-QED}

\title{Learning to Classify Threading in Melts of Rings}

\newcommand*{\affEDI}{School of Physics and Astronomy, University of Edinburgh, Peter Guthrie Tait Road, Edinburgh, EH9 3FD, UK}
\newcommand*{\affGEN}{MRC Human Genetics Unit, Institute of Genetics and Cancer, University of Edinburgh, Edinburgh EH4 2XU, UK}
\newcommand*{\affVIE}{Faculty of Physics, University of Vienna, Boltzmanngasse 5, 1090 Vienna, Austria}
\newcommand*{\affCAM}{Yusuf Hamied Department of Chemistry, University of Cambridge, Lensfield Road, Cambridge CB2 1EW, UK}

\author{Filippo Conforto}
\affiliation{\affEDI}
\author{Roman Staňo}
\affiliation{\affCAM}
\author{Jan Smrek}
\affiliation{\affVIE}
\author{Davide Michieletto}
\email{davide.michieletto@ed.ac.uk}
\affiliation{\affEDI}
\affiliation{\affGEN}

\date{\today}
\begin{abstract}
Dense melts of nonconcatenated ring polymers exhibit anomalous dynamics and nonlinear rheology, in some cases linked to long-lived inter-ring threading entanglement. Detecting these geometric constraints remains computationally demanding, with different methods being based on different geometric features tuned to specific models. Here, we introduce a machine-learning framework that detects threading directly from the writhe of ring conformations and we demonstrate its broad generalizability. 
Trained on simulations of isolated, distance-constrained pairs of short rings, a one-dimensional convolutional neural network (CNN1D) operating on the writhe profile classifies threading states with over 98\% accuracy on held-out ring pairs. Remarkably, this model generalizes without retraining to equilibrium dense melts spanning a wide range of densities and chain lengths ($N = 100$ to $1600$), maintaining true-positive rates above 91\% with false-positive rates near zero. Misclassified conformations are limited to physically ambiguous, shallow threading events which are arguably not impacting the dynamics of the rings. Effectively, writhe captures the interaction between the two rings, allowing writhe-based classifiers to improve the topological analysis of large ring-polymer systems with a 3-fold speed-up over the state-of-the-art minimal-surface detection. These results establish the writhe-trained neural networks as an accurate, scalable tool for characterizing entanglement in ring polymer melts.
\end{abstract}
\maketitle

\section{\label{sec:introduction} Introduction}

Dense solutions and melts of long nonconcatenated, unknotted ring polymers \cite{Tubiana2024a, Schroeder2025a, Kruteva2023a, FrankKamenetskii1975a} display dynamical and rheological behaviour that is fundamentally incompatible with the standard tube and reptation paradigms developed for entangled linear chains~\cite{Doi1986a}. 
In these systems, rings remain globally compact and non-Gaussian, and feature self-similar dynamics with power-law stress relaxation, none of these features observed for polymers with trivial linear topology \cite{Kapnistos2008a, Doi2015a, Pasquino2013a, Vettorel2009a, Halverson2011a, Halverson2011b, Halverson2012a}.
One of the most puzzling aspects of ring melts is the role of inter-ring threading -- a long-lived entanglement in which one ring passes through the opening of another ring \cite{Smrek2016a, Tsalikis2016a}.
On one hand, microscopic theories based on self-similar relaxation \cite{Ge2016a, Smrek2015a} or soft caging dynamics \cite{Tu2023a, Mei2024a} completely ignore the threading and yet almost quantitatively reproduce the experimental results in equilibrium.
On the other hand, molecular simulations \cite{Michieletto2017a, Michieletto2016a, Michieletto2014a, Smrek2019a} clearly show that threading is not a secondary detail but the central microscopic mechanism underpinning rheological anomalies, generating persistent directed constraint networks between the rings, driving dynamical arrest in diverse conditions~\cite{Lo2013a, Lee2019a, Smrek2020a, OConnor2020a, Huang2019a, Borger2020a, Stano2026a}.
Moreover, threading of rings by linear chains in ring/linear blends is the key mechanism governing the relaxation but also miscibility of these systems \cite{Hagita2021a, Grest2023a, Vigil2024a, Parisi2021a, Kim2026a}, while a similar behavior for pure ring melts is still only speculated.
Reconciling these contradiction and understanding the formation and dynamics of inter-ring threading is therefore central to any microscopic theory of ring polymer rheology.

However, detecting threading in solutions or melts of ring polymers is challenging. The act of threading one ring through another does not change the global topology of the system and is therefore not associated to a change of topological invariants, such as the Gauss linking number. For this reason, more sophisticated algorithms have been developed to detect threadings in melts of rings. Among the most notable, leveraging a background scaffold to define changes in \emph{locally defined} topological invariants~\cite{Michieletto2014a}, probing the indirect signatures in dynamic coupling and cooperative motion \cite{Lee2015a, Michieletto2021a}, primitive path reduction \cite{Tsalikis2014a}, constructing minimal surfaces spanning the rings and detecting intersection of such surfaces~\cite{Smrek2016a, Lang2013a}, and exploiting topological data analysis and parameter-free persistent homology diagrams to define ``homological'' threadings~\cite{Landuzzi2020a}. These studies have independently discovered that threadings are abundant in dense and entangled solutions of rings and number of entanglements scales linearly with the rings' length. Additionally, they demonstrated that threading plays a central role in the rings' slowdown and the resultant rheological behaviour.

Despite this recent progress, reliable detection of threadings in melts of rings remains technically challenging and computationally demanding. 
For example, the methods based on primitive path reduction \cite{Tsalikis2014a} have to alter the native configuration to detect a threading, in process losing fine information about the depth of the threading or about the position of the piercing.
The methods based on the minimal surface of the ring \cite{Smrek2016a, Lang2013a} do not interfere with the configuration, but constructing the surface of each ring needs its own minimization simulation and subsequent localization of piercings between every pair of rings. 
There is therefore a need for new, scalable and general tools that can detect, classify, and track threadings directly from polymer configurations, and that can generalise to different chain length or polymer models. 
Beyond the development of new threading detection algorithms, there is an interesting polymer physics question to be addressed: are there specific geometric signatures of threading configurations? Indeed, inspired by recent discoveries on polymer knots~\cite{Sleiman2023a, Sleiman2022a, Braghetto2023a, Vandans2020a}, one may argue that specific patterns in how polymers ``writhe'' around themselves and each other may reflect specific types of entanglements, among which, threading.

In this work, we address this question by developing a machine learning framework for the detection and characterisation of threading in melts of ring polymers. Specifically, we construct descriptors of ring conformations and mutual penetrations, and train supervised models to detect threaded rings without explicitly computing minimal surfaces or homological invariants. Our approach is inspired by recent successes of geometric learning in knot topology, where we demonstrated that deep neural networks trained using a ``writhe density'' descriptor could accurately classify knot topology and localise knotted segments in fluctuating polymers~\cite{Sleiman2023a}. By adapting these ideas to the context of ring-threading, we show that patterns in the mutual writhe density field can serve as effective ``read-outs'' of underlying threading constraints, enabling fast and robust detection across a broad range of chain lengths, concentrations, and stiffnesses.

Most remarkably, whilst we train our ML on simulations of pairs of distance-constrained short ring polymers, we show that our model can accurately identify threadings in melts of polymers and can generalise to different chain lengths and melt densities that had not seen in the training. We argue that this generalisation is here possible thanks to the fact that threadings are ``local'' features that mostly occur between loops shorter than the entanglement length~\cite{Ubertini2022a}. Finally, we demonstrate that our ML classification algorithm is 3-fold faster than measuring intersections of minimal surfaces, i.e. the most accurate threading detection algorithm currently available, a result that encourages large scale analysis of threadings in polymer melts.

\section{\label{sec:methods} Model and Methods}

\subsection{Polymer model}

To represent an unknotted ring polymer, we used a standard coarse-grained bead-spring model \cite{Kremer1990a} in which every pair of monomeric units at distance $x$ interact through the purely repulsive Weeks-Chandler-Anderson (WCA) potential
\begin{equation}
    U_{\mathrm{WCA}}(x) = 4\epsilon \left[ \left( \dfrac{\sigma}{x} \right)^{12} - \left( \dfrac{\sigma}{x} \right)^6 + \dfrac{1}{4} \right] \cdot H \left( 2^{1/6} - \dfrac{x}{\sigma} \right)
\end{equation}
where $\sigma$ and $\epsilon = \kT = 1/\beta$ set the length scale and energy scale respectively and $H(\cdot)$ is the Heavside step-function. To emulate the bonds between monomeric units, we used the finite extensible nonlinear elastic (FENE) potential
\begin{equation}
    U_{\mathrm{FENE}}(x) = -\dfrac{K_{\mathrm{FENE}}}{2} R_0^2 \ln \left[ 1 - \left( \dfrac{x}{R_0} \right)^2 \right]
\end{equation}
where $R_0 = 1.5\sigma$ is the maximal bond length beyond which the potential diverges and $K_{\mathrm{FENE}} = 30 \kT$, which in combination with the WCA potential yields bonds with equilibrium length $\approx 0.97\sigma$. Unless stated otherwise, we did not use any angle bending potential and polymers were fully flexible.

\subsection{Training set generation}

To create the training set, we simulated a pair of unknotted and unlinked rings with $N = 100$ monomeric units each in a cubic simulation box with $L = 150 \sigma$ in the periodic boundary conditions, corresponding to a dilute solution (see Fig.~\ref{fig:fig1}). To make sure we sampled the configurations at every possible ring separation, we applied a biasing potential
\begin{equation}
    U_{\mathrm{bias}}(r) = \dfrac{K_{\mathrm{bias}}}{2} (r - r_0)^2
\end{equation}
between the centres of mass of the two rings at a distance $r$, with $K_\mathrm{bias} = 2.5 \kT/\sigma^2$ and $r_0$ spanning the interval $[0\sigma,25 \sigma]$, covering the inter-ring separation up to four radii of gyration of a single ring $R_g \approx 5.5 \sigma$.

To solve this model, we used the Molecular dynamics with Langevin thermostat as implemented in LAMMPS~\cite{Plimpton1995a} (\texttt{29Sep2021}), where we numerically integrated the equation of motion $m\ddot{\mathbf{x}}_i(t) = \mathbf{F}_i - \gamma m \dot{\mathbf{x}}_i(t) + \mathbf{Y}_i(t)$, for each of $i$ particles with position $\mathbf{x}_i$, while $\langle Y_i(t) \rangle = 0$ and $\langle Y_i(t) Y_j(t') \rangle = 2 \gamma m \kT \delta_{ij} \delta(t - t')$ for each of the Cartesian coordinates of the random force $\mathbf{Y}_i(t)$. The force $\mathbf{F}_i$ originated from the gradients of the WCA and FENE potentials and also from the biasing potential applied to the centres of mass of the rings. Finally, the mass was arbitrarily set to $m = 1$ hence defining the time scale $\tau = 1/\gamma = (m \sigma^2 / \kT)^{1/2} = 200 \delta t$, where $\delta t = 0.005 \tau$ is the time step of the Velocity-Verlet integrator. For each of 40 values of sampled $r_0$, we ran 16 independent simulations with different initial configurations and different random seeds, each lasting $10^6 \tau$, greatly exceeding the Rouse time. All in all, we obtained $\sim 5 \cdot 10^5$ independent configurations of a pair of rings to be used for the training, testing and validation.

\subsubsection*{Minimal surface construction}
\label{sec:methods:minsurf}

For every sampled configuration, we constructed the minimal surface for each of the rings using the Surface Evolver~\cite{Brakke1992a}, where we first spanned an initial triangular mesh over the ring, using both the immobilized monomeric units and added virtual particles as the mesh vertices. The latter were iteratively evolved under the effect of the surface tension of the mesh tiles to find the surface with the minimal energy. During the minimization, the initial disc-like topology of the surface was maintained, but the triangulation was progressively adapted to weed out the small triangles to drive the surface to the Delaunay triangulation. The procedure is identical to the one used in our previous publications~\cite{Smrek2016a, Smrek2019a, Stano2023a}, and representative configurations of threaded rings are shown in \reffig{fig:fig1}(A,B).

Once the minimal surfaces were obtained, we checked if any of the bond vectors of a ring intersected the surface of the other ring; when such crossing was found we would classify that as a threading. Because the rings are not topologically linked, each time we found a segment crossing the minimal surface in one direction, we found another segment of the first ring crossing the minimal surface in the opposite direction.  
A threading ring can thus be understood as (at least) two countour lengths divided by the a minimal surface of the passively threaded one (see \reffig{fig:fig1}(C,D)) 
To further quantify the threading, we counted the number of the monomers on either side of the minimal surface of the threaded ring, and defined this the minimum of either of these two numbers as the separation length $L_{\mathrm{sep}}$.

In order to label each pair configuration we computed the threading by using the minimal surface construction. In Fig.~\ref{fig:fig1}C we show that as a function of distance, the pairs of rings can display single, mutual or no threading; and that the probability of each state is dependent on the separation of the rings. Additionally, we can also measure the threading depth (see Fig.~\ref{fig:fig1}D) and show that it decreases with ring-ring separation. 

\subsection{Testing set generation}

The testing set consisted of concentrated solutions of ring polymers made of $M = 512$ unlinked unknotted polymers of $N = 100$ at a range of monomer densities $\rho\sigma^3$ in the interval $[0.01, 0.50]$ corresponding to $[0.07\rho^*, 3.48\rho^*]$ in terms of overlap density, here defined as $\rho^* = 3N / (4 \pi R_g^3)$, hence covering the range from dilute, through semi-dilute to concentrated solutions; representative snapshots at low and high density are shown in \reffig{fig:fig3}(A) and \reffig{fig:fig3}(B), respectively. To prepare such a system, we first initialized the rings as well separated unknots on a primitive cubic lattice and then slowly compressed the box by iterative rescaling of the coordinates while running the Langevin dynamics to reach the box length $L$ yielding the desired density. At this density, we equilibrated the system for $2 \cdot 10^5\tau$ followed by a production run of $10^6\tau$ entailing many Rouse times at each of the densities ($\sim 2500\tau$ in the densest system). 

Finally, to probe generalization to longer chains, we also consider melts of $M = 200$ unlinked and unknotted polymers with length $N = 400$, $N = 800$ and $N = 1600$ at monomer density $\rho \sigma^3 = 0.85$ identical to the systems studied in Ref.~\cite{Halverson2011a, Halverson2011b}. 
For this test case, the rings were not fully flexible, and a soft angle bending potential
\begin{equation}
    U_{\mathrm{bend}}(\theta) = K_{\mathrm{bend}} (1 - \cos(\theta - \pi))
\end{equation}
was applied on angle $\theta$ defined by every pair of the consecutive bond vectors, and $K_{\mathrm{bend}} = 1.5 \kT$. The initial configurations were prepared on circular paths on lattice and then collapsed by a short simulation with an added attractive intra-chain potential, a procedure detailed in Ref.~\cite{Vettorel2010a}. 
After removing the attractive potential and adjusting the density of the melt by isobaric rescaling, the system was equilibrated and simulated for more than $10^7 \tau$, covering several Rouse times.

\subsection{Neural Network Architectures}

To predict threading from polymer configurations, we tested three distinct neural network architectures (implemented in TensorFlow/Keras\cite{tensorflow2015-whitepaper}). All architectures were designed to output a binary threading classification task, i.e. threaded vs unthreaded. 

\subsubsection*{Recurrent Neural Network (RNN)}
The RNN architecture was designed as an input layer followed by an optional batch-normalization layer, a unidirectional long short-term memory (LSTM) layer (100 units, returning full sequences), a bidirectional LSTM layer (100 units per direction), and a subsequent unidirectional LSTM layer (100 units). The network ends with a final LSTM layer (100 units) that returns a categorical output, following the approach of Sleiman et al.\cite{Sleiman2023a}. All recurrent layers used a softmax activation function and zero recurrent dropout.

\subsubsection*{Two-dimensional Convolutional Neural Network (CNN2D)} The CNN2D architecture, intended for matrix-valued inputs (e.g., writhe density matrices, see below), consists of an optional batch-normalization layer, a 2D convolutional layer (32 filters, $3 \times 3$ kernel), a $2 \times 2$ max-pooling layer, and a second 2D convolutional layer (32 filters, $3 \times 3$ kernel). A final global average-pooling layer enables the network to process inputs of variable spatial dimension natively. This was followed by a dense layer of 80 units. All the layers used a softmax activation function.

\subsubsection*{One-dimensional Convolutional Neural Network (CNN1D)}
The CNN1D architecture processed sequential inputs through three 1D convolutional layers (32, 32, and 64 filters, respectively; kernel width 10; ``same'' padding), followed by a global max-pooling layer. Global max-pooling was used to allow variable-length sequences. A dense layer of 80 units preceded the output. All the layers used a softmax activation function.

\subsection{Output Layers, Loss Functions, and Optimization}

For the classification task, the output layer used a softmax activation, while the model was trained by minimizing the sparse categorical cross-entropy loss,
\begin{equation}
    L_{\mathrm{CE}}
    = -\frac{1}{N} \sum_{i=1}^{N} \log p_{i,c_i} ,
\end{equation}
where $p_{i,c_i}$ is the predicted probability assigned to the true integer-encoded class $c_i$ of sample $i$. All models were compiled with a configurable optimizer (e.g., Adam), and classification accuracy was monitored as an auxiliary metric during training.

\begin{figure*}[t!]
\centering
\includegraphics[scale = 0.25]{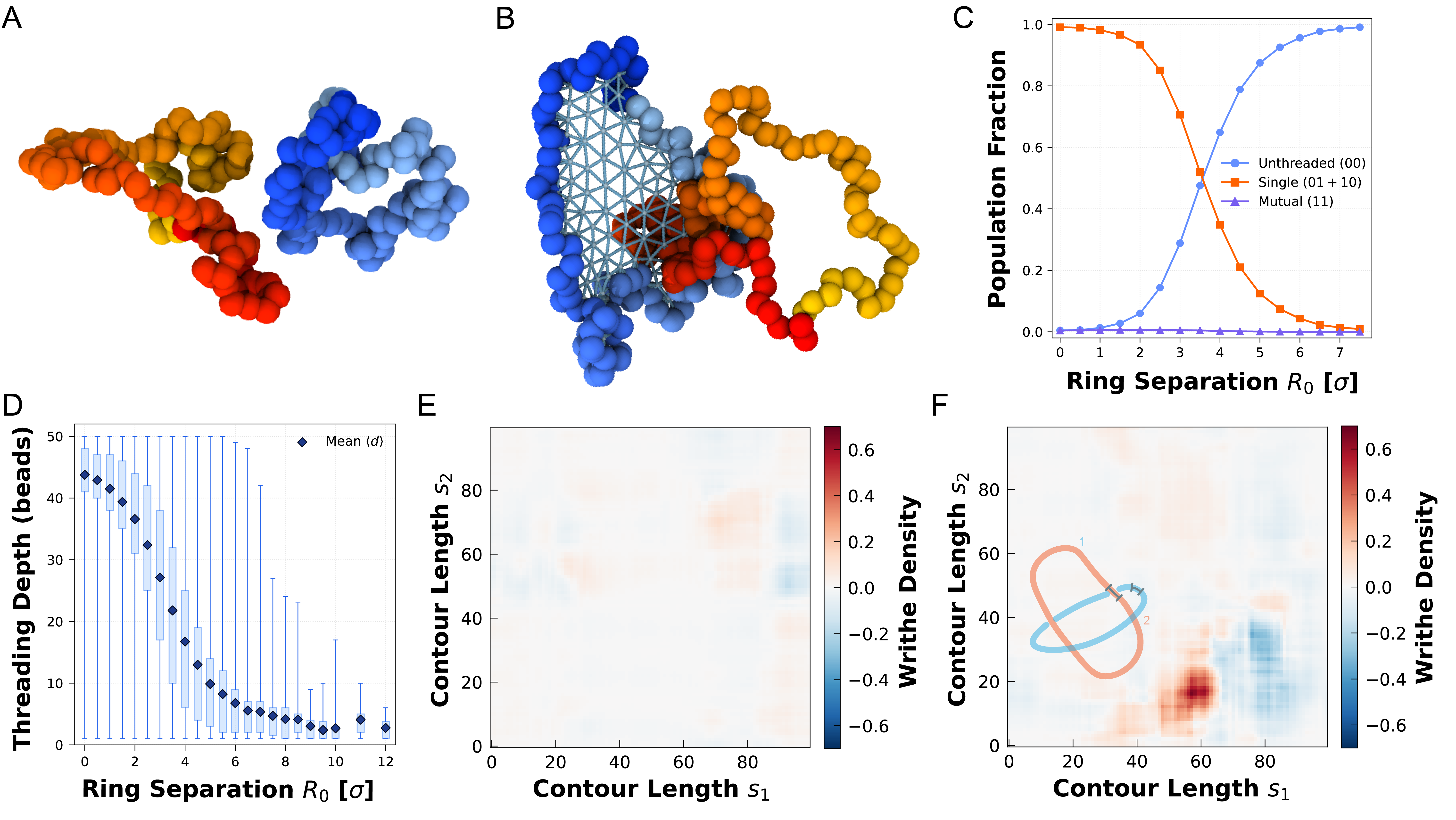}
\caption{\textbf{Extracting local features of inter-ring threading from writhe density maps.} \textbf{(A-B)} A pair of (A) non-threading and (B) threading rings from the dataset generated in Ref.~\cite{Stano2023a}. Minimal surface (grey mesh) is shown spanning the passively threaded ring. \textbf{(C)} Fraction of non-, singly and mutually threaded configurations as a function of inter-ring separation. \textbf{(D)} Threading depth distribution as a function of inter-ring separation. The box limits represent the lower and upper quartile of the distribution, the whisker limits show the minimum and maximum, while the dark point is the mean of the distribution.
\textbf{(E)} Writhe density map for the unthreaded configuration in (A). \textbf{(F)} Writhe density map for the threaded configuration in (B).
}
\label{fig:fig1}
\end{figure*}

\subsection{Data Preparation}
\label{sec:methods:writhe}

\begin{figure*}[t!]
\centering
\includegraphics[scale = 0.25]{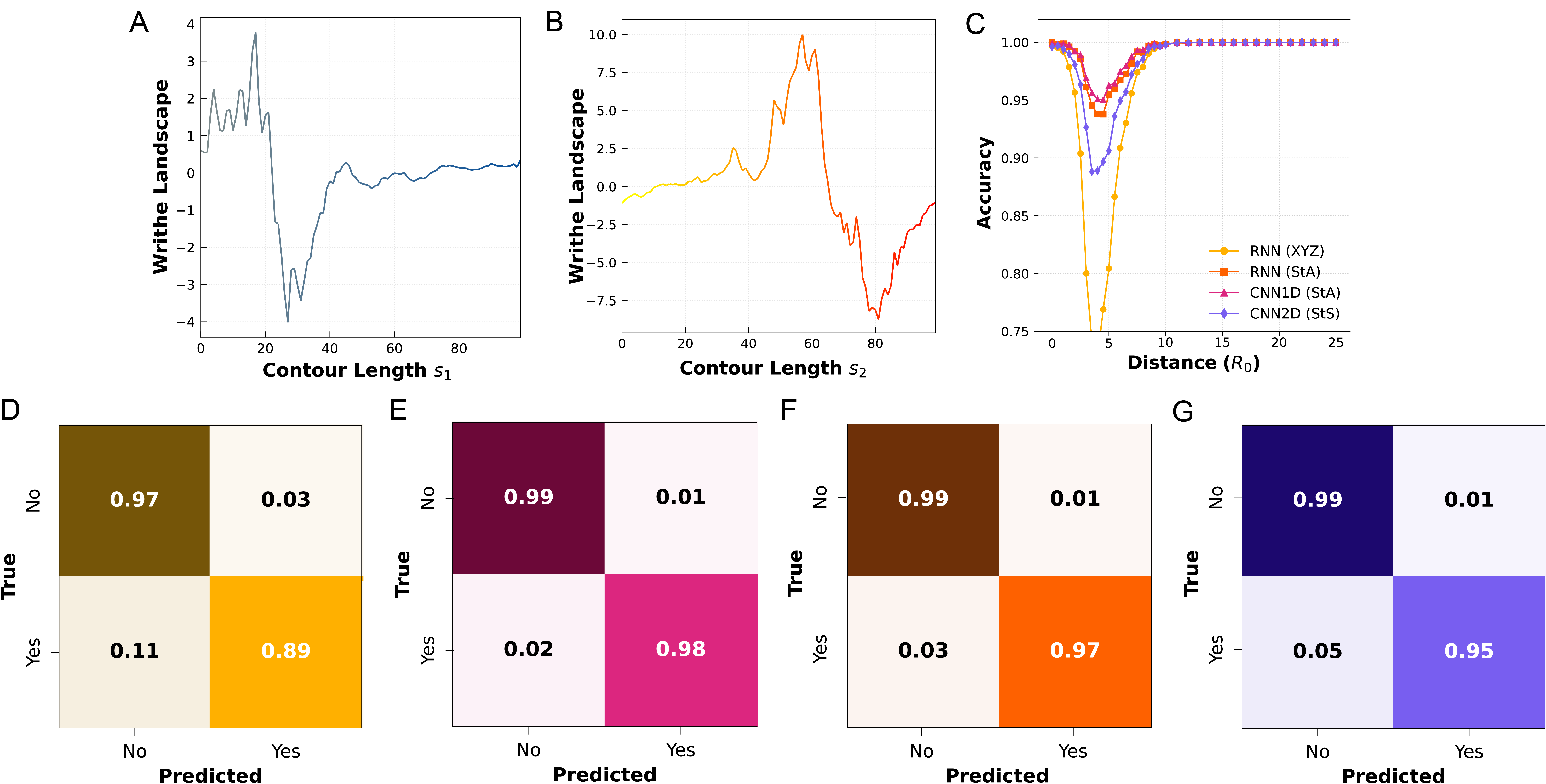}
\caption{\textbf{Writhe-based threading classification outperforms XYZ.} \textbf{(A-B)} Writhe landscapes $w_1(s)$ and $w_2(s)$ for the two threading rings shown in Fig.~\ref{fig:fig1}B.  
\textbf{(C)} Accuracy of classification compared across different model architectures and for different training features (XYZ vs writhe landscape ``StA'' and writhe density ``StS'') as a function of separation between the rings in the held-out data.
\textbf{(D-G)} From left to right: confusion matrices for the models RNN (XYZ), RNN (writhe landscape), CNN1D (writhe landscape) and CNN2D (writhe density). }
\label{fig:fig2}
\end{figure*}

The first feature used for training is \texttt{XYZ}, the raw Cartesian coordinates of the two chains ordered along their contours.  From these coordinates we construct the mutual writhe of the two rings. For two closed curves $\gamma_1$ and $\gamma_2$, the Gauss double integral
\begin{equation}
W = \frac{1}{4\pi} \oint_{\gamma_1}\!\oint_{\gamma_2}
\frac{\left( \mathrm{d}\mathbf{r}_1 \times \mathrm{d}\mathbf{r}_2 \right)
\cdot \left( \mathbf{r}_1 - \mathbf{r}_2 \right)}
{\left| \mathbf{r}_1 - \mathbf{r}_2 \right|^3}
\label{eq:gauss}
\end{equation}
measures the total signed winding of one curve about the other. Rather than evaluating the scalar $W$ (which would be zero for any ring pair), we retain the \emph{local} contribution to the integrand: for the pair of segments at contour positions $s_1$ (on ring~1) and $s_2$ (on ring~2) we compute the corresponding integrand of Eq.~\eqref{eq:gauss} directly from the \texttt{XYZ} coordinates, in turn obtaining a signed value $\omega(s_1, s_2)$. Collecting these values gives the two-dimensional signed writhe density matrix, shown in Fig.~\ref{fig:fig1}(E). From $\omega(s_1, s_2)$ we can compute the signed writhe landscape  representation by summing the writhe density map over rows (or columns), in turn collapsing the matrix into a one dimensional function that tracks the total writhe experienced by a single segment against the entirety of the partner ring,
\begin{equation}
w_1(s_1) = \sum_{s_2} \omega(s_1, s_2),
\qquad
w_2(s_2) = \sum_{s_1} \omega(s_1, s_2), \, .
\end{equation}
The two profiles $w_1(s_1)$ and $w_2(s_2)$ are concatenated to form the input feature. An example of $w_2$ for the configuration of Fig.~\ref{fig:fig1}(B) is shown in Fig.~\ref{fig:fig2}(A,B).

Threading labels were derived from the minimal-surface analysis of Sec.~\ref{sec:methods}. A binary threading label was assigned to each ring by testing whether any of its segments pierced the partner's minimal surface, while the threading depth was defined as the separation length $L_{\mathrm{sep}}$ introduced above.

\subsection{Training/Validation/Test Split and Evaluation} 

The dataset was partitioned into training (70\%), test, and validation subsets using a fixed random seed (42) to ensure reproducibility. Specifically, an initial $70/30$ split separated the training set from the remainder, and the remaining 30\% was further split ($70/30$) into test and validation sets. The threading-depth values were partitioned using an identical sequence of splits and random seed, guaranteeing exact correspondence between the classification labels and their associated depth values in the test set.

For heavily imbalanced datasets, like the ones we are dealing here where there are many more non-threading conformations than threading ones, it is useful to consider the F1 score to measure accuracy as it combines both precision $\mathrm{P} = \frac{\mathrm{TP}}{\mathrm{TP}+\mathrm{FP}}$, and recall $\mathrm{R} = \frac{\mathrm{TP}}{\mathrm{TP}+\mathrm{FN}}$ (where TP = true positives, FP = false positives, FN = false negatives).
The F1 score is the harmonic mean of precision and recall, and is defined as
\begin{equation}
F1 = 2 \frac{\mathrm{P}\times\mathrm{R}}
{\mathrm{P}+\mathrm{R}}
= \frac{2\,\mathrm{TP}}{2\,\mathrm{TP}+\mathrm{FP}+\mathrm{FN}} \, .
\end{equation}

\section{\label{sec:results} Results and Discussion}

\subsection*{Writhe-trained ML models outperform XYZ-trained ones on threading classification in isolated and distance-constrained pairs of rings.}

We first trained the networks on a dataset composed of isolated pairs of rings generated in Ref.~\cite{Stano2023a}. The training set spanned configurations from deeply interpenetrating (threaded) to well separated (unthreaded) (see Fig.~\ref{fig:fig1}E,F and SI). Each pair was labelled by explicitly constructing a minimal surface spanning one ring and testing whether the contour of the partner ring passed through it. If it did, we would label the pair as threading, assign passive/active threading labels, and measure the threading depth, as done in Ref.~\cite{Smrek2016a} (see Sec.~\ref{sec:methods:minsurf}). This procedure yielded the ground truth for our system.

Next, we asked if the writhe density map could provide information on the entanglement between the rings. For each pair of rings, we thus computed the writhe density map as described in Sec.~\ref{sec:methods:writhe}. As expected, these maps yield clear features when rings are threaded as opposed to a flatter 2D landscape when rings are not threading, akin to what observed in our previous calculations on knotted and self-entangled polymers~\cite{Michieletto2016softmatter,Sleiman2022a}.  Importantly, the map's features strongly correlate with regions in which polymer segments intertwine with each other (Fig.~\ref{fig:fig1}(E,F)). The writhe density map effectively tracks local crossings between the two polymers and their relative crossing sign. Specifically, positive and negative peaks in the writhe density map appear localised at the contour positions where the actively threading ring enters and exists the minimal surface of the passively threaded ring (see Fig.~\ref{fig:fig1}(B,F)). These features encode not only for the sign-sensitive local winding associated with a threading event, but also for the information related to the threading depth. 

This qualitative observation motivated us to explore the use of writhe density map as input feature for ML models. More specifically, we expect the information required to identify threading to be concentrated in a small number of localised signals rather than distributed diffusely across the configuration, a structure that convolutional filters are well suited to detect. With this in mind, we trained both recurrent neural networks (RNN) and convolutional neural networks (CNN) in 2D over writhe density or in 1D over, the writhe landscape, the one-dimensional writhe profile obtained by summing the writhe density map over its rows and columns (Fig.~\ref{fig:fig2}(A-B)).

\begin{figure*}[t!]
\centering
\includegraphics[scale = 0.25]{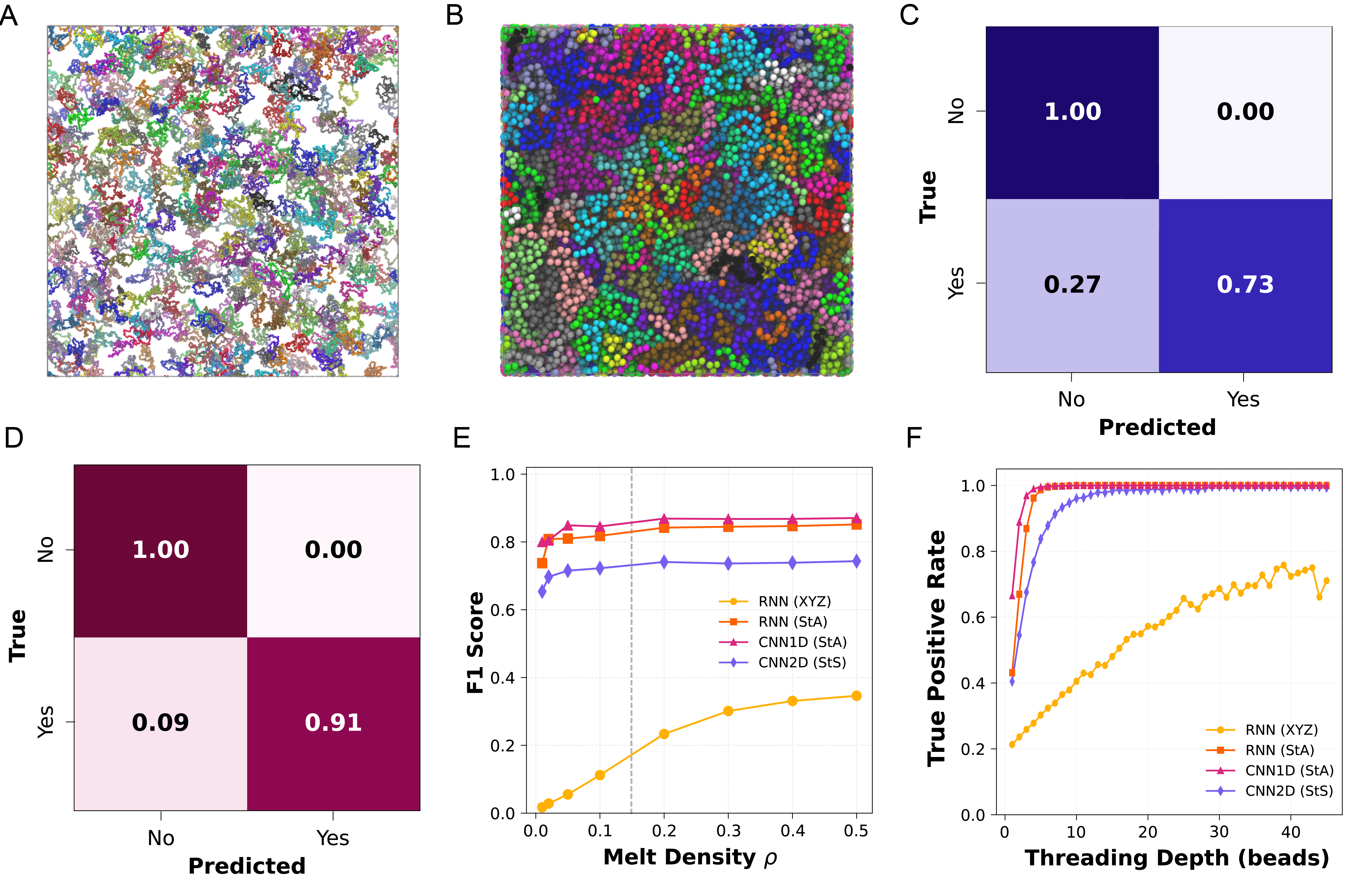}
\caption{\textbf{ML models generalise to classify threadings in the dense solutions.} \textbf{(A-B)} Snapshot of a solution of $N=100$ rings at density (A) $\rho=0.05\,\sigma^{-3}$ and (B) $\rho=0.5\,\sigma^{-3}$ . \textbf{(C-D)} Confusion matrices for the classification of threadings between pairs of rings in the melt (ring length $N=100$): (A) CNN2D model on writhe density, (B) CNN1D model on writhe landscape. \textbf{(E)} F1 score as a function of the monomer density across the ML models considered in this work. \textbf{(F)} Recall (True Positives) as a function of depth of threading. 
}
\label{fig:fig3}
\end{figure*}

Both convolutional architectures classify the held-out test set of ring pairs with high fidelity. The CNN2D, which operates directly on the two-dimensional writhe density map, achieves a true-negative rate of $0.99$ and a true-positive rate of $0.95$ (Fig.~\ref{fig:fig2}(E)). Remarkably, the CNN1D performs even better, reaching a true-negative rate of $0.99$ and a true-positive rate of $0.98$ (Fig.~\ref{fig:fig2}(F)), despite taking as input only the writhe landscape (which should contain less information than the writhe density). This is a significant practical result: the row/column reduction discards no information relevant to threading detection, so the more compact writhe landscape representation not only matches but slightly exceeds the accuracy of the full matrix while allowing for a substantially smaller and cheaper model. Figure~\ref{fig:fig2}(C) breaks down the classification accuracy performance as a function of the separation between the rings. All models are essentially 100\% accurate at large separations, where the rings cannot thread, and exhibit a dip in accuracy at intermediate distances of a $R_g$, where some rings thread and some don't. At short distances, most rings are threaded and therefore all models return to be highly accurate. The writhe-based convolutional models (CNN1D and CNN2D fed with writhe landscape/writhe density) retain accuracies above $94\%$ even in the most challenging regime, whereas the RNN operating on raw Cartesian coordinates (XYZ) degrades markedly, confirming that the writhe representation supplies a far more discriminative input than the bare geometry.

\subsection*{ML models trained on isolated rings can generalise and accurately classify threadings in dense solutions.}

Having validated the models on isolated pairs of distance-constrained rings, we next asked whether a network trained exclusively on distance-constrained, isolated pairs of rings can generalise to the far more challenging environment of a dense solutions. In these systems, each ring can be simultaneously threaded by many neighbours and typically assumes conformations that are different from the ones seen in isolated rings, particularly in dense conditions when rings are long \cite{Halverson2011a}. To test this, we ask the pre-trained ML models to classify threadings between pairs of rings (with length $N=100$) extracted from systems between 0.01 and 0.5 $\sigma^{-3}$ monomer densities (see snapshots in Fig.~\ref{fig:fig3}(A,B)). We discovered that the CNN2D correctly predicts only $73\%$ of true threading events (Fig.~\ref{fig:fig3}(C)), and is outperformed by the CNN1D which correctly identifies $91\%$ and essentially zero false-positives (Fig.~\ref{fig:fig3}(D)).

\begin{figure*}[t!]
\centering
\includegraphics[scale = 0.25]{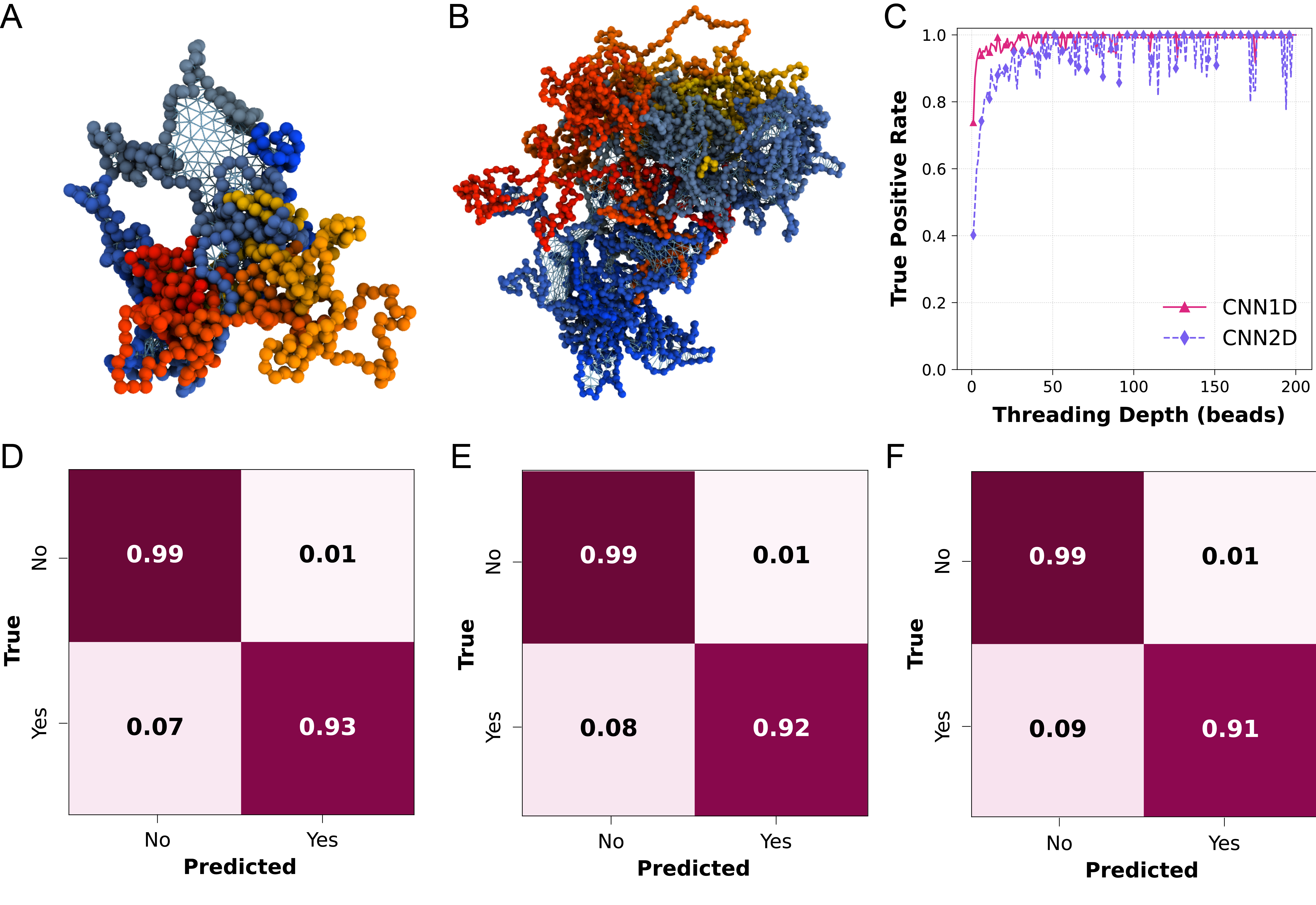}
\caption{\textbf{Write-trained CNN models generalise to longer rings and melts.} \textbf{(A-B)} Snapshots of selected subsets of rings from the melts, with (A) $N=400$ and (B) $N=1600$. \textbf{(C)} Recall (true positives) as a function of threading depth for $N=400$ for CNN1D and CNN2D. \textbf{(D-F)} Confusion matrices reporting the threading classification of pairs of rings in the melt via the CNN1D model: (A) $N=400$, (B) $N=800$ (C) $N=1600$.
}
\label{fig:fig4}
\end{figure*}

The superior generalisability of the CNN1D is consistent with the compactness of the writhe landscape input: by projecting the pairwise structure of the writhe density, the writhe landscape is less sensitive to the small scale fluctuations due to many-body interactions and generalises more robustly to dense solutions. Figure~\ref{fig:fig3}(E) shows that this behaviour is preserved across densities: the F1 score of both, the CNN1D and RNN trained on the writhe landscape remains high and essentially flat from the dilute limit up to the maximum density considered, $\rho = 0.5\,\sigma^{-3}$. The CNN2D trained on the writhe density behaves in a similar way but is systematically lower for the same number of model parameters. Finally, the RNN trained on raw XYZ coordinates performs poorly throughout.

The origin of the residual misclassification is clarified in Fig.~\ref{fig:fig3}(F), where we report the accurate classification rate as a function of threading depth. The writhe-based models reach near-perfect recall for threadings that are deeper than a handful of beads; most of the misclassified conformations are concentrated at very shallow threading depths. These shallow threadings are marginal, short-lived, and physically ambiguous even under the explicit minimal-surface definition, so this shortfall reflects the intrinsic fuzziness of the definition of a threading in the limit of shallow depth.

\subsection*{Write-trained ML models generalise to different chain lengths and polymer melts.}

Finally, we tested the most demanding form of generalisation: applying the CNN1D, still trained only on pairs of short rings, to melts ($\rho = 0.85$ $\sigma^{-3}$) of longer chains, specifically $N=400$, $800$, and $1600$ with non-zero bending stiffness. Surprisingly, the CNN1D model maintains excellent performance across the range of ring lengths, with true-positive rates of $0.93$, $0.92$, and $0.91$, respectively, and a false-positive rate of just $0.01$ across the systems (Fig.~\ref{fig:fig4}(D)-(F)). 

The fact that these confusion matrices score essentially the same across chain lengths much longer than the one in the training set ($N=100$) demonstrates that the model has learned a genuinely local signature of threading rather than a feature tied to a particular ring size or contour length. As before, the depth-resolved recall, or true positives rate (Fig.~\ref{fig:fig4}(C)) confirms that the few remaining errors are confined to the shallowest threads, with essentially perfect detection once the thread penetrates more than a small number of beads.

Beyond accuracy, the classifiers offer a substantial reduction in computational cost. On a single CPU core, the minimal-surface construction that defines the ground truth required roughly $200$ minutes for one $N=400$ snapshot and about $711$ minutes $N=800$ beads long (each snapshot having 200 rings). The writhe-based ML pipeline instead processes the same snapshot in three stages: a distance filter first rejects the $\sim\!74\%$ of ring pairs too far apart to thread ($d \ge 2.5\,R_g$) at $5\,\mu\mathrm{s}$ per pair, in under $0.1\,\mathrm{s}$ per snapshot. For the $\sim\!5{,}200$ surviving candidates the writhe density map is then evaluated in $163\,\mathrm{ms}$ per pair and the CNN1D classifies each in a further $4\,\mathrm{ms}$, giving a single-core snapshot time of $\approx 14.5$~minutes at $N=400$ and $\approx 56$~minutes at $N=800$, with a $\sim 3\times$ speed-up compared to the minimal surface. Crucially, this budget is taken almost entirely by the calculation of the writhe density, which accounts for $\gtrsim 97\%$ of the runtime, whereas the network inference contributes only $\sim\!20\,\mathrm{s}$ ($\sim\!2\%$). Both stages parallelise readily, the writhe density evaluation scales to $49\,\mathrm{ms}$ per pair on $64$ cores and GPU inference reduces the network cost to below a second, bringing the full snapshot to $\approx 4.2$~minutes at $N=400$. In terms of complexity, both methods scale comparably as $\mathcal{O}(M N^{2})$ or $\mathcal{O}(M^{2} N^{2})$ without the distance filter, where $M$ is the number of rings in the system. The bottleneck for the minimal-surface is the intersection verification routine, where position of each bond is checked against that of each triangle of the minimal surface. As the rings in the melt are tree-like \cite{Smrek2016a}, the surface and the number of triangles scales with $N$ and the cost of the minimization procedure itself is largely subdominant for the investigated cases. However, for an open ring conformation the minimal surface area is expected to scale as $R_{\rm g}^{2}\sim N^{2\nu}$, with $\nu>1/2$, giving contribution of $\mathcal{O}(N^{1+2\nu})$. Therefore, a potential advantage of the classifier lies in the robust $N^{2}$ complexity of writhe landscape calculation irrespective of the conformation. 
High-throughput threading analysis of large ring melts or other systems is thus feasible on modest resources and scales straightforwardly as more compute power becomes available.
\section{\label{sec:conclusions} Conclusions}

In summary, we have shown that (i) the writhe density and writhe landscape are mathematical objects that display local features that correlate with threading events; (ii) compact convolutional models, and in particular a one-dimensional CNN (CNN1D) acting on the reduced writhe landscape profile can classify isolated ring pairs with $\gtrsim 98\%$ accuracy; (iii) these models transfer without retraining to dense solutions and melts across a wide range of densities and (iv) chain lengths, with residual errors confined to physically ambiguous shallow threads. Finally, ML improve the classification speed of the state-of-the-art tool for detecting threadings, i.e. the explicit calculation of minimal-surfaces. Our results establish writhe-trained NNs as an accurate and scalable tool for characterizing non-trivial topological structures in polymeric systems.


\section{\label{sec:sm} Supplementary Material}
Supplementary material contains supporting analysis showing the effect of the input representation and further analysis for the writhe recognition.

\begin{acknowledgments}
DM and FC thank the Royal Society for support through a University Research Fellowship. This project has received funding from the European Research Council (ERC) under the European Union's Horizon 2020 research and innovation program (grant agreement No 947918, TAP). 
DM also acknowledges the Leverhulme Trust for funding. 
RS is funded by the U.K. Research and Innovation (UKRI) Engineering and Physical Sciences Research Council (EPSRC) under the U.K. Government’s guarantee scheme (EP/Z002028/1), following funding by the European Research Council (Consolidator Grant) under the European Union’s Horizon Europe research and innovation programme.
JS thanks his friends and colleagues for many interesting and entertaining physics discussions throughout the years.  

\end{acknowledgments}

\section*{Data Availability statement}
The data that support the findings of this study are openly available in Zenodo at https://doi.org/10.5281/zenodo.22893849, Ref. ~\onlinecite{zenodo_models}. The code that supports the findings of this study is openly available in GitLab at https://git.ecdf.ed.ac.uk/s2469797/ml-threadings-project.



\bibliography{aipsamp}

@PREAMBLE{
 "\providecommand{\noopsort}[1]{}" 
 # "\providecommand{\singleletter}[1]{#1}%" 
}

@article{Michieletto2016softmatter,
  author  = {Michieletto, D.},
  title   = {On the tree-like structure of rings in dense solutions},
  journal = {Soft Matter},
  year    = {2016},
  volume  = {12},
  issue   = {47},
  pages   = {9485--9500},
  doi     = {10.1039/c6sm02168a},
  url     = {https://doi.org/10.1039/c6sm02168a},
  publisher = {The Royal Society of Chemistry}
}

@article{Borger2020a,
  author = {Borger, Anine and Wang, Wendi and O'Connor, Thomas C. and Ge, Ting and Grest, Gary S. and Jensen, Grethe V. and Ahn, Junyoung and Chang, Taihyun and Hassager, Ole and Mortensen, Kell and Vlassopoulos, Dimitris and Huang, Qian},
  doi = {10.1021/acsmacrolett.0c00607},
  issn = {2161-1653, 2161-1653},
  journal = {ACS Macro Lett.},
  number = {10},
  pages = {1452--1457},
  title = {Threading--Unthreading Transition of Linear-Ring Polymer Blends in Extensional Flow},
  url = {https://pubs.acs.org/doi/10.1021/acsmacrolett.0c00607},
  volume = {9},
  year = {2020},
}

@article{Braghetto2023a,
  title = {Machine Learning Understands Knotted Polymers},
  volume = {56},
  ISSN = {1520-5835},
  url = {http://dx.doi.org/10.1021/acs.macromol.2c02555},
  DOI = {10.1021/acs.macromol.2c02555},
  number = {7},
  journal = {Macromolecules},
  publisher = {American Chemical Society (ACS)},
  author = {Braghetto,  Anna and Kundu,  Sumanta and Baiesi,  Marco and Orlandini,  Enzo},
  year = {2023},
  month = Mar,
  pages = {2899–2909}
}

@article{Brakke1992a,
  title = {The Surface Evolver},
  volume = {1},
  ISSN = {1944-950X},
  url = {http://dx.doi.org/10.1080/10586458.1992.10504253},
  DOI = {10.1080/10586458.1992.10504253},
  number = {2},
  journal = {Experimental Mathematics},
  publisher = {Informa UK Limited},
  author = {Brakke,  Kenneth A.},
  year = {1992},
  month = Jan,
  pages = {141–165}
}

@book{Doi1986a,
  title     = {The Theory of Polymer Dynamics},
  author    = {Doi, Masao and Edwards, Sam F.},
  year      = {1986},
  publisher = {Clarendon Press},
  address   = {Oxford},
  isbn      = {9780198520337}
}

@article{Doi2015a,
  title={Melt Rheology of Ring Polystyrenes with Ultrahigh Purity},
  volume={48},
  url={https://pubs.acs.org/doi/10.1021/acs.macromol.5b00076},
  DOI={10.1021/acs.macromol.5b00076},
  number={9},
  journal={Macromolecules},
  author={Doi, Yuya and Matsubara, Kazuki and Ohta, Yutaka and Nakano, Tomohiro and Kawaguchi, Daisuke and Takahashi, Yoshiaki and Takano, Atsushi and Matsushita, Yushu},
  year={2015},
  pages={3140–3147},
}

@article{FrankKamenetskii1975a,
  author={Frank-Kamenetskii, M. D. and Lukashin, A. V. and Vologodskii, A. V.},
  title={Statistical mechanics and topology of polymer chains},
  journal={Nature},
  year={1975},
  day={01},
  volume={258},
  number={5534},
  pages={398-402},
  issn={1476-4687},
  doi={10.1038/258398a0},
  url={https://doi.org/10.1038/258398a0}
}

@article{Ge2016a,
  author = {Ge, Ting and Panyukov, Sergey and Rubinstein, Michael},
  title = {Self-Similar Conformations and Dynamics in Entangled Melts and Solutions of Nonconcatenated Ring Polymers},
  journal = {Macromolecules},
  volume = {49},
  number = {2},
  pages = {708-722},
  year = {2016},
  doi = {10.1021/acs.macromol.5b02319},
  URL = {https://doi.org/10.1021/acs.macromol.5b02319},
}

@article{Grest2023a,
  author = {Grest, Gary S. and Ge, Ting and Plimpton, Steven J. and Rubinstein, Michael and O'Connor, Thomas C.},
  doi = {10.1021/acspolymersau.2c00050},
  issn = {2694-2453, 2694-2453},
  journal = {ACS Polym. Au},
  number = {2},
  pages = {209--216},
  title = {Entropic Mixing of Ring/Linear Polymer Blends},
  url = {https://pubs.acs.org/doi/10.1021/acspolymersau.2c00050},
  volume = {3},
  year = {2023},
}

@article{Hagita2021a,
  title = {Molecular Dynamics Simulations of Ring Shapes on a Ring Fraction in Ring–Linear Polymer Blends},
  volume = {54},
  ISSN = {1520-5835},
  url = {http://dx.doi.org/10.1021/acs.macromol.1c00656},
  DOI = {10.1021/acs.macromol.1c00656},
  number = {17},
  journal = {Macromolecules},
  publisher = {American Chemical Society (ACS)},
  author = {Hagita,  Katsumi and Murashima,  Takahiro},
  year = {2021},
  month = Aug,
  pages = {8043–8051}
}

@article{Halverson2011a,
  Author = {Halverson, Jonathan D. and Lee, Won Bo and Grest, Gary S. and Grosberg, Alexander Y. and Kremer, Kurt},
  Doi = {10.1063/1.3587137},
  Journal = {J. Chem. Phys.},
  Number = {20},
  Pages = {204904},
  Title = {{Molecular dynamics simulation study of nonconcatenated ring polymers in a melt. I. Statics}},
  Url = {https://doi.org/10.1063/1.3587137},
  Volume = {134},
  Year = {2011},
}

@article{Halverson2011b,
  Author = {Halverson, Jonathan D. and Lee, Won Bo and Grest, Gary S. and Grosberg, Alexander Y. and Kremer, Kurt},
  Doi = {10.1063/1.3587138},
  Journal = {J. Chem. Phys.},
  Number = {20},
  Pages = {204905},
  Title = {{Molecular dynamics simulation study of nonconcatenated ring polymers in a melt. II. Dynamics}},
  Url = {https://doi.org/10.1063/1.3587138},
  Volume = {134},
  Year = {2011},
}

@article{Halverson2012a,
  title = {Rheology of Ring Polymer Melts: From Linear Contaminants to Ring-Linear Blends},
  volume = {108},
  ISSN = {1079-7114},
  url = {http://dx.doi.org/10.1103/PhysRevLett.108.038301},
  DOI = {10.1103/physrevlett.108.038301},
  number = {3},
  journal = {Physical Review Letters},
  publisher = {American Physical Society (APS)},
  author = {Halverson,  Jonathan D. and Grest,  Gary S. and Grosberg,  Alexander Y. and Kremer,  Kurt},
  year = {2012},
  month = Jan 
}

@article{Huang2019a,
  title = {Unexpected Stretching of Entangled Ring Macromolecules},
  volume = {122},
  ISSN = {1079-7114},
  url = {http://dx.doi.org/10.1103/PhysRevLett.122.208001},
  DOI = {10.1103/physrevlett.122.208001},
  number = {20},
  journal = {Physical Review Letters},
  publisher = {American Physical Society (APS)},
  author = {Huang,  Q. and Ahn,  J. and Parisi,  D. and Chang,  T. and Hassager,  O. and Panyukov,  S. and Rubinstein,  M. and Vlassopoulos,  D.},
  year = {2019},
  month = May 
}

@article{Kapnistos2008a,
  doi = {10.1038/nmat2292},
  url = {https://doi.org/10.1038/nmat2292},
  year = {2008},
  publisher = {Springer Science and Business Media {LLC}},
  volume = {7},
  number = {12},
  pages = {997--1002},
  author = {M. Kapnistos and M. Lang and D. Vlassopoulos and W. Pyckhout-Hintzen and D. Richter and D. Cho and T. Chang and M. Rubinstein},
  title = {Unexpected power-law stress relaxation of entangled ring polymers},
  journal = {Nat. Mater.}
}

@article{Kim2026a,
  title = {Direct Observation of Topologically Enhanced Interfacial Miscibility in Ring–Linear Polymers},
  ISSN = {2374-7951},
  url = {http://dx.doi.org/10.1021/acscentsci.6c00828},
  DOI = {10.1021/acscentsci.6c00828},
  journal = {ACS Central Science},
  publisher = {American Chemical Society (ACS)},
  author = {Kim,  Seong Eun and Kim,  Dogyun and Park,  Se Hwa and Kong,  Tae Yeon and Lee,  June Hyuk and Kim,  Kyoung Taek and Kim,  So Youn},
  year = {2026},
  month = Sept 
}

@article{Kremer1990a,
  title = {Dynamics of entangled linear polymer melts: A molecular-dynamics simulation},
  volume = {92},
  ISSN = {1089-7690},
  url = {http://dx.doi.org/10.1063/1.458541},
  DOI = {10.1063/1.458541},
  number = {8},
  journal = {The Journal of Chemical Physics},
  publisher = {AIP Publishing},
  author = {Kremer,  Kurt and Grest,  Gary S.},
  year = {1990},
  month = Apr,
  pages = {5057–5086}
}

@article{Kruteva2023a,
  title = {Topology Matters: Conformation and Microscopic Dynamics of Ring Polymers},
  volume = {56},
  ISSN = {1520-5835},
  url = {http://dx.doi.org/10.1021/acs.macromol.3c00560},
  DOI = {10.1021/acs.macromol.3c00560},
  number = {18},
  journal = {Macromolecules},
  publisher = {American Chemical Society (ACS)},
  author = {Kruteva,  Margarita and Allgaier,  J\"{u}rgen and Richter,  Dieter},
  year = {2023},
  month = Sept,
  pages = {7203–7229}
}

@article{Landuzzi2020a,
  title = {Persistence homology of entangled rings},
  volume = {2},
  ISSN = {2643-1564},
  url = {http://dx.doi.org/10.1103/PhysRevResearch.2.033529},
  DOI = {10.1103/physrevresearch.2.033529},
  number = {3},
  journal = {Physical Review Research},
  publisher = {American Physical Society (APS)},
  author = {Landuzzi,  Fabio and Nakamura,  Takenobu and Michieletto,  Davide and Sakaue,  Takahiro},
  year = {2020},
  month = Sept 
}

@article{Lang2013a,
  title = {Ring Conformations in Bidisperse Blends of Ring Polymers},
  volume = {46},
  ISSN = {1520-5835},
  url = {http://dx.doi.org/10.1021/ma301359b},
  DOI = {10.1021/ma301359b},
  number = {3},
  journal = {Macromolecules},
  publisher = {American Chemical Society (ACS)},
  author = {Lang,  M.},
  year = {2013},
  month = Jan,
  pages = {1158–1166}
}

@article{Lee2015a,
  title = {Slowing Down of Ring Polymer Diffusion Caused by Inter‐Ring Threading},
  volume = {36},
  ISSN = {1521-3927},
  url = {http://dx.doi.org/10.1002/marc.201400713},
  DOI = {10.1002/marc.201400713},
  number = {11},
  journal = {Macromolecular Rapid Communications},
  publisher = {Wiley},
  author = {Lee,  Eunsang and Kim,  Soree and Jung,  YounJoon},
  year = {2015},
  month = Apr,
  pages = {1115–1121}
}

@article{Lee2019a,
  title = {Slow Dynamics of Ring Polymer Melts by Asymmetric Interaction of Threading Configuration: Monte Carlo Study of a Dynamically Constrained Lattice Model},
  volume = {11},
  ISSN = {2073-4360},
  url = {http://dx.doi.org/10.3390/polym11030516},
  DOI = {10.3390/polym11030516},
  number = {3},
  journal = {Polymers},
  publisher = {MDPI AG},
  author = {Lee,  Eunsang and Jung,  YounJoon},
  year = {2019},
  month = Mar,
  pages = {516}
}

@article{Lo2013a,
  doi = {10.1209/0295-5075/102/58005},
  url = {https://doi.org/10.1209/0295-5075/102/58005},
  year = {2013},
  publisher = {{IOP} Publishing},
  volume = {102},
  number = {5},
  pages = {58005},
  author = {Wei-Chang Lo and Matthew S. Turner},
  title = {The topological glass in ring polymers},
  journal = {EPL},
}

@article{Mei2024a,
  title = {Unified understanding of the impact of semiflexibility,  concentration,  and molecular weight on macromolecular-scale ring diffusion},
  volume = {121},
  ISSN = {1091-6490},
  url = {http://dx.doi.org/10.1073/pnas.2403964121},
  DOI = {10.1073/pnas.2403964121},
  number = {31},
  journal = {Proceedings of the National Academy of Sciences},
  publisher = {National Academy of Sciences},
  author = {Mei,  Baicheng and Grest,  Gary S. and Liu,  Songyue and O’Connor,  Thomas C. and Schweizer,  Kenneth S.},
  year = {2024},
  month = July 
}

@article{Michieletto2014a,
  title = {Threading Dynamics of Ring Polymers in a Gel},
  volume = {3},
  ISSN = {2161-1653},
  url = {http://dx.doi.org/10.1021/mz500060c},
  DOI = {10.1021/mz500060c},
  number = {3},
  journal = {ACS Macro Letters},
  publisher = {American Chemical Society (ACS)},
  author = {Michieletto,  Davide and Marenduzzo,  Davide and Orlandini,  Enzo and Alexander,  Gareth P. and Turner,  Matthew S.},
  year = {2014},
  month = Feb,
  pages = {255–259}
}

@article{Michieletto2016a,
  author = {Michieletto, Davide  and Turner, Matthew S.},
  title = {{A topologically driven glass in ring polymers}},
  journal = {Proc. Natl. Acad. Sci. U.S.A.},
  volume = {113},
  number = {19},
  pages = {5195-5200},
  year = {2016},
  doi = {10.1073/pnas.1520665113},
  URL = {https://www.pnas.org/doi/abs/10.1073/pnas.1520665113},
}

@article{Michieletto2017a,
  title = {Glassiness and Heterogeneous Dynamics in Dense Solutions of Ring Polymers},
  author = {Michieletto, Davide and Nahali, Negar and Rosa, Angelo},
  journal = {Phys. Rev. Lett.},
  volume = {119},
  issue = {19},
  pages = {197801},
  year = {2017},
  publisher = {American Physical Society},
  doi = {10.1103/PhysRevLett.119.197801},
  url = {https://link.aps.org/doi/10.1103/PhysRevLett.119.197801}
}

@article{Michieletto2021a,
  author = {Michieletto, Davide and Sakaue, Takahiro},
  title = {{Dynamical Entanglement and Cooperative Dynamics in Entangled Solutions of Ring and Linear Polymers}},
  journal = {ACS Macro Lett.},
  volume = {10},
  number = {1},
  pages = {129-134},
  year = {2021},
  doi = {10.1021/acsmacrolett.0c00551},
  URL = {https://doi.org/10.1021/acsmacrolett.0c00551},
}

@article{OConnor2020a,
  title = {Topological Linking Drives Anomalous Thickening of Ring Polymers in Weak Extensional Flows},
  author = {O'Connor, Thomas C. and Ge, Ting and Rubinstein, Michael and Grest, Gary S.},
  journal = {Phys. Rev. Lett.},
  volume = {124},
  issue = {2},
  pages = {027801},
  year = {2020},
  publisher = {American Physical Society},
  doi = {10.1103/PhysRevLett.124.027801},
  url = {https://link.aps.org/doi/10.1103/PhysRevLett.124.027801}
}

@article{Parisi2021a,
  author = {Parisi, Daniele  and Kaliva, Maria  and Costanzo, Salvatore  and Huang, Qian  and Lutz, Pierre J.  and Ahn, Junyoung  and Chang, Taihyun  and Rubinstein, Michael  and Vlassopoulos, Dimitris},
  title = {Nonlinear rheometry of entangled polymeric rings and ring-linear blends},
  journal = {J. Rheol.},
  volume = {65},
  number = {4},
  pages = {695-711},
  year = {2021},
  doi = {10.1122/8.0000186},
  URL = {https://doi.org/10.1122/8.0000186},
}

@article{Pasquino2013a,
  author = {Pasquino, Rossana and Vasilakopoulos, Thodoris C. and Jeong, Youn Cheol and Lee, Hyojoon and Rogers, Simon and Sakellariou, George and Allgaier, J{\"u}rgen and Takano, Atsushi and Br{\'a}s, Ana R. and Chang, Taihyun and Goo{\ss}en, Sebastian and Pyckhout-Hintzen, Wim and Wischnewski, Andreas and Hadjichristidis, Nikos and Richter, Dieter and Rubinstein, Michael and Vlassopoulos, Dimitris},
  doi = {10.1021/mz400344e},
  issn = {2161-1653, 2161-1653},
  journal = {ACS Macro Lett.},
  number = {10},
  pages = {874--878},
  title = {Viscosity of Ring Polymer Melts},
  url = {https://pubs.acs.org/doi/10.1021/mz400344e},
  volume = {2},
  year = {2013},
}

@article{Plimpton1995a,
  title = {Fast Parallel Algorithms for Short-Range Molecular Dynamics},
  journal = {J. Comput. Phys.},
  volume = {117},
  number = {1},
  pages = {1-19},
  year = {1995},
  issn = {0021-9991},
  author = {Steve Plimpton},
}

@article{Schroeder2025a,
    author = {Schroeder, Charles M. and Everaers, Ralf and Kremer, Kurt and Kruteva, Margarita and Likos, Christos N. and McKenna, Gregory B. and O’Connor, Thomas and Ravi Prakash, J. and Richter, Dieter and Robertson-Anderson, Rae and Rubinstein, Michael and Schweizer, Kenneth S. and Vlassopoulos, Dimitris},
    title = {Ring polymer physics and rheology: Challenges and opportunitiesa)},
    journal = {Journal of Rheology},
    volume = {70},
    number = {1},
    pages = {183-216},
    year = {2025},
    month = {12},
    issn = {0148-6055},
    doi = {10.1122/8.0001099},
    url = {https://doi.org/10.1122/8.0001099},
}

@article{Sleiman2022a,
  title = {Geometric Predictors of Knotted and Linked Arcs},
  volume = {2},
  ISSN = {2694-2453},
  url = {http://dx.doi.org/10.1021/acspolymersau.2c00021},
  DOI = {10.1021/acspolymersau.2c00021},
  number = {5},
  journal = {ACS Polymers Au},
  publisher = {American Chemical Society (ACS)},
  author = {Sleiman,  Joseph L. and Burton,  Robin H. and Caraglio,  Michele and Gutierrez Fosado,  Yair Augusto and Michieletto,  Davide},
  year = {2022},
  month = July,
  pages = {341–350}
}

@article{Sleiman2023a,
  title = {Geometric learning of knot topology},
  volume = {20},
  ISSN = {1744-6848},
  url = {http://dx.doi.org/10.1039/d3sm01199b},
  DOI = {10.1039/d3sm01199b},
  number = {1},
  journal = {Soft Matter},
  publisher = {Royal Society of Chemistry (RSC)},
  author = {Sleiman,  Joseph Lahoud and Conforto,  Filippo and Fosado,  Yair Augusto Gutierrez and Michieletto,  Davide},
  year = {2024},
  pages = {71–78}
}

@ARTICLE{Smrek2015a,
  doi = {10.1088/0953-8984/27/6/064117},
  url = {https://doi.org/10.1088/0953-8984/27/6/064117},
  year = {2015},
  publisher = {{IOP} Publishing},
  volume = {27},
  number = {6},
  pages = {064117},
  author = {Jan Smrek and Alexander Y Grosberg},
  title = {Understanding the dynamics of rings in the melt in terms of the annealed tree model},
  journal = {J. Phys. Condens. Matter},
}

@article{Smrek2016a,
  title = {Minimal Surfaces on Unconcatenated Polymer Rings in Melt},
  volume = {5},
  ISSN = {2161-1653},
  url = {http://dx.doi.org/10.1021/acsmacrolett.6b00289},
  DOI = {10.1021/acsmacrolett.6b00289},
  number = {6},
  journal = {ACS Macro Letters},
  publisher = {American Chemical Society (ACS)},
  author = {Smrek,  Jan and Grosberg,  Alexander Y.},
  year = {2016},
  month = Jun,
  pages = {750–754}
}

@article{Smrek2019a,
  title = {Threading of Unconcatenated Ring Polymers at High Concentrations: Double-Folded vs Time-Equilibrated Structures},
  volume = {8},
  ISSN = {2161-1653},
  url = {http://dx.doi.org/10.1021/acsmacrolett.8b00828},
  DOI = {10.1021/acsmacrolett.8b00828},
  number = {2},
  journal = {ACS Macro Letters},
  publisher = {American Chemical Society (ACS)},
  author = {Smrek,  Jan and Kremer,  Kurt and Rosa,  Angelo},
  year = {2019},
  month = Jan,
  pages = {155–160}
}

@article{Smrek2020a,
  title = {Active topological glass},
  volume = {11},
  ISSN = {2041-1723},
  url = {http://dx.doi.org/10.1038/s41467-019-13696-z},
  DOI = {10.1038/s41467-019-13696-z},
  number = {1},
  journal = {Nature Communications},
  publisher = {Springer Science and Business Media LLC},
  author = {Smrek,  Jan and Chubak,  Iurii and Likos,  Christos N. and Kremer,  Kurt},
  year = {2020},
  month = Jan 
}

@article{Stano2023a,
  title = {To thread or not to thread? Effective potentials and threading interactions between asymmetric ring polymers},
  volume = {19},
  ISSN = {1744-6848},
  url = {http://dx.doi.org/10.1039/D2SM01177H},
  DOI = {10.1039/d2sm01177h},
  number = {1},
  journal = {Soft Matter},
  publisher = {Royal Society of Chemistry (RSC)},
  author = {Staňo,  Roman and Likos,  Christos N. and Smrek,  Jan},
  year = {2023},
  pages = {17–30}
}

@article{Stano2026a,
  title = {Actively Induced Supercoiling Can Slow Down Plasmid Solutions by Trapping the Threading Entanglements},
  volume = {20},
  ISSN = {1936-086X},
  url = {http://dx.doi.org/10.1021/acsnano.5c10811},
  DOI = {10.1021/acsnano.5c10811},
  number = {21},
  journal = {ACS Nano},
  publisher = {American Chemical Society (ACS)},
  author = {Staňo,  Roman and Rusková,  Renáta and Račko,  Dušan and Smrek,  Jan},
  year = {2026},
  month = May,
  pages = {14998–15007}
}

@article{Tsalikis2014a,
  title = {Threading of Ring Poly(ethylene oxide) Molecules by Linear Chains in the Melt},
  volume = {3},
  ISSN = {2161-1653},
  url = {http://dx.doi.org/10.1021/mz5002096},
  DOI = {10.1021/mz5002096},
  number = {8},
  journal = {ACS Macro Letters},
  publisher = {American Chemical Society (ACS)},
  author = {Tsalikis,  Dimitrios G. and Mavrantzas,  Vlasis G.},
  year = {2014},
  month = July,
  pages = {763–766}
}

@article{Tsalikis2016a,
  author = {Tsalikis, Dimitrios G. and Mavrantzas, Vlasis G. and Vlassopoulos, Dimitris},
  title = {Analysis of Slow Modes in Ring Polymers: Threading of Rings Controls Long-Time Relaxation},
  journal = {ACS Macro Lett.},
  volume = {5},
  number = {6},
  pages = {755-760},
  year = {2016},
  doi = {10.1021/acsmacrolett.6b00259},
  URL = {https://doi.org/10.1021/acsmacrolett.6b00259},
  eprint = {https://doi.org/10.1021/acsmacrolett.6b00259}
}

@article{Tu2023a,
  author = {Tu, Michael Q. and Davydovich, Oleg and Mei, Baicheng and Singh, Piyush K. and Grest, Gary S. and Schweizer, Kenneth S. and O’Connor, Thomas C. and Schroeder, Charles M.},
  title = {Unexpected Slow Relaxation Dynamics in Pure Ring Polymers Arise from Intermolecular Interactions},
  journal = {ACS Polym. Au},
  volume = {3},
  number = {4},
  pages = {307-317},
  year = {2023},
  doi = {10.1021/acspolymersau.2c00069},
  URL = {https://doi.org/10.1021/acspolymersau.2c00069},
}

@article{Tubiana2024a,
  title = {Topology in soft and biological matter},
  journal = {Physics Reports},
  volume = {1075},
  pages = {1-137},
  year = {2024},
  issn = {0370-1573},
  doi = {https://doi.org/10.1016/j.physrep.2024.04.002},
  url = {https://www.sciencedirect.com/science/article/pii/S0370157324001339},
  author = {Luca Tubiana and Gareth P. Alexander and Agnese Barbensi and Dorothy Buck and Julyan H.E. Cartwright and Mateusz Chwastyk and Marek Cieplak and Ivan Coluzza and Simon Čopar and David J. Craik and Marco {Di Stefano} and Ralf Everaers and Patrícia F.N. Faísca and Franco Ferrari and Achille Giacometti and Dimos Goundaroulis and Ellinor Haglund and Ya-Ming Hou and Nevena Ilieva and Sophie E. Jackson and Aleksandre Japaridze and Noam Kaplan and Alexander R. Klotz and Hongbin Li and Christos N. Likos and Emanuele Locatelli and Teresa López-León and Thomas Machon and Cristian Micheletti and Davide Michieletto and Antti Niemi and Wanda Niemyska and Szymon Niewieczerzal and Francesco Nitti and Enzo Orlandini and Samuela Pasquali and Agata P. Perlinska and Rudolf Podgornik and Raffaello Potestio and Nicola M. Pugno and Miha Ravnik and Renzo Ricca and Christian M. Rohwer and Angelo Rosa and Jan Smrek and Anton Souslov and Andrzej Stasiak and Danièle Steer and Joanna Sułkowska and Piotr Sułkowski and De Witt L. Sumners and Carsten Svaneborg and Piotr Szymczak and Thomas Tarenzi and Rui Travasso and Peter Virnau and Dimitris Vlassopoulos and Primož Ziherl and Slobodan Žumer},
}

@article{Ubertini2022a,
  author = {Ubertini, Mattia Alberto and Smrek, Jan and Rosa, Angelo},
  title = {Entanglement Length Scale Separates Threading from Branching of Unknotted and Non-concatenated Ring Polymers in Melts},
  journal = {Macromolecules},
  volume = {55},
  number = {23},
  pages = {10723-10736},
  year = {2022},
  doi = {10.1021/acs.macromol.2c01264},
  URL = {https://doi.org/10.1021/acs.macromol.2c01264},
}

@article{Vandans2020a,
  title = {Identifying knot types of polymer conformations by machine learning},
  volume = {101},
  ISSN = {2470-0053},
  url = {http://dx.doi.org/10.1103/PhysRevE.101.022502},
  DOI = {10.1103/physreve.101.022502},
  number = {2},
  journal = {Physical Review E},
  publisher = {American Physical Society (APS)},
  author = {Vandans,  Olafs and Yang,  Kaiyuan and Wu,  Zhongtao and Dai,  Liang},
  year = {2020},
  month = Feb 
}

@article{Vettorel2009a,
  title = {Statistics of polymer rings in the melt: a numerical simulation study},
  volume = {6},
  ISSN = {1478-3975},
  url = {http://dx.doi.org/10.1088/1478-3975/6/2/025013},
  DOI = {10.1088/1478-3975/6/2/025013},
  number = {2},
  journal = {Physical Biology},
  publisher = {IOP Publishing},
  author = {Vettorel,  Thomas and Grosberg,  Alexander Y and Kremer,  Kurt},
  year = {2009},
  month = July,
  pages = {025013}
}

@article{Vettorel2010a,
  title = {Development of Entanglements in a Fully Disentangled Polymer Melt},
  volume = {19},
  ISSN = {1521-3919},
  url = {http://dx.doi.org/10.1002/mats.200900065},
  DOI = {10.1002/mats.200900065},
  number = {1},
  journal = {Macromolecular Theory and Simulations},
  publisher = {Wiley},
  author = {Vettorel,  Thomas and Kremer,  Kurt},
  year = {2010},
  month = Feb,
  pages = {44–56}
}

@article{Vigil2024a,
  title = {Measuring Topological Constraint Relaxation in Ring-Linear Polymer Blends},
  volume = {133},
  ISSN = {1079-7114},
  url = {http://dx.doi.org/10.1103/PhysRevLett.133.118101},
  DOI = {10.1103/physrevlett.133.118101},
  number = {11},
  journal = {Physical Review Letters},
  publisher = {American Physical Society (APS)},
  author = {Vigil,  Daniel L. and Ge,  Ting and Rubinstein,  Michael and O’Connor,  Thomas C. and Grest,  Gary S.},
  year = {2024},
  month = Sept 
}

@misc{zenodo_models,
  doi = {10.5281/ZENODO.22893849},
  url = {https://zenodo.org/doi/10.5281/zenodo.22893849},
  author = {Conforto,  Filippo and Staňo,  Roman},
  title = {Pre-trained Deep Learning Models for Detecting Topological Threadings in Ring Polymers},
  publisher = {Zenodo},
  year = {2026},
  copyright = {Creative Commons Attribution 4.0 International}
}

@misc{tensorflow2015-whitepaper,
title={ {TensorFlow}: Large-Scale Machine Learning on Heterogeneous Systems},
url={https://www.tensorflow.org/},
note={Software available from tensorflow.org},
author={
    Mart\'{i}n~Abadi and
    Ashish~Agarwal and
    Paul~Barham and
    Eugene~Brevdo and
    Zhifeng~Chen and
    Craig~Citro and
    Greg~S.~Corrado and
    Andy~Davis and
    Jeffrey~Dean and
    Matthieu~Devin and
    Sanjay~Ghemawat and
    Ian~Goodfellow and
    Andrew~Harp and
    Geoffrey~Irving and
    Michael~Isard and
    Yangqing Jia and
    Rafal~Jozefowicz and
    Lukasz~Kaiser and
    Manjunath~Kudlur and
    Josh~Levenberg and
    Dandelion~Man\'{e} and
    Rajat~Monga and
    Sherry~Moore and
    Derek~Murray and
    Chris~Olah and
    Mike~Schuster and
    Jonathon~Shlens and
    Benoit~Steiner and
    Ilya~Sutskever and
    Kunal~Talwar and
    Paul~Tucker and
    Vincent~Vanhoucke and
    Vijay~Vasudevan and
    Fernanda~Vi\'{e}gas and
    Oriol~Vinyals and
    Pete~Warden and
    Martin~Wattenberg and
    Martin~Wicke and
    Yuan~Yu and
    Xiaoqiang~Zheng},
  year={2015},
}

\end{document}